\documentclass{llncs}
\usepackage{epsfig,footnote,makeidx}

\newcommand{\BEGINITEM}{\begin{list}{$\bullet$}{\itemsep 0in \parsep 0in}}
\newcommand{\ENDITEM}{\end{list}}
\newcommand{\BEGINDASHITEM}{\begin{list}{---}{itemsep 0in \parsep 0in}}
\newcommand{\ENDDASHITEM}{\end{list}}
\newcommand{\BEGINNUMITEM}{\begin{list}{\arabic{numList-count})}{
\itemsep 0in \parsep 0in \usecounter{numList-count}}}
\newcommand{\ENDNUMITEM}{\end{list}}
\newcommand{\BEGINNUMDOTITEM}{\begin{list}{\arabic{numList-count}.}{
\itemsep 0in \parsep 0in \usecounter{numList-count}}}
\newcommand{\ENDNUMDOTITEM}{\end{list}}

\newcommand{\BEGINTNUMDOTITEM}{\begin{list}{\arabic{numList-count}.}{
\topsep 0in \footnotesep 0in \leftmargin 0.2in \itemsep 0in \parsep 0in \usecounter{numList-count}}}
\newcommand{\ENDTNUMDOTITEM}{\end{list}}

{\begin{displaymath}\begin{array}{l}} %
{\end{array}\end{displaymath}}

\newcounter{numList-count}
\newcounter{algorithm-count}
\makesavenoteenv{table} 
\begin{document}
\frontmatter          
\pagestyle{empty}
\title{Formal Modeling in a Commercial Setting: A Case Study}

\author{Andre Wong \and Marsha Chechik}
\institute{Department of Computer Science,
University of Toronto,\\ Toronto, ON M5S 3G4, Canada,\\
\email{\{andre,chechik\}@cs.toronto.edu},\\ WWW home page:
\texttt{http://www.cs.toronto.edu/\homedir chechik}}

\maketitle


\begin{abstract}
This paper describes a case study conducted in collaboration with Nortel to
demonstrate the feasibility of applying formal modeling techniques 
to telecommunication systems.  A formal description language, SDL, was chosen
by our qualitative CASE tool evaluation to model a multimedia-messaging system
described by an 80-page natural language specification. Our model was used to
identify errors in the software requirements document and to derive test
suites, shadowing the existing development process and keeping track of a
variety of productivity data.
%
\end{abstract}

\section{Introduction}
\label{intro}
For a long time, researchers and practitioners have been seeking ways
to improve productivity in the software development process.  Precise
documentation of software specifications has been advocated as one of
the viable approaches~\cite{parnas93a}.  If high quality
specifications are crucial to the success of system developments, it
seems logical to apply rigorous specification techniques to the
requirements for ensuring their completeness and consistency.

The majority of successful applications of formal modeling have been
confined to safety-critical
projects~\cite{crow96,heimdahl96a,joannou93} where software
correctness is the pivotal goal. In contrast, the commercial software
industry seeks practical techniques that can be seamlessly integrated
into the existing development processes and improve productivity;
absolute quality is often a desirable but not crucial
objective. Although the feasibility of formal specifications has been
demonstrated in commercial settings~\cite{hall96,hoare95,mataga95},
the overall adoption of the idea has been slow.  Most companies, such
as the Canadian-based telecommunications company
Nortel\footnote{Nortel, for the purpose of this paper, refers to the
Toronto Multimedia Applications Center of Nortel Networks.}, opt to
rely on manual inspections of natural-language specifications as the
only technique to detect errors in them, even though the results
have been suboptimal. If the advantages of better quality
specifications, such as a better understanding of the system and less
error-prone designs, do not provide an adequate justification,
more benefits can be obtained by leveraging the investment in the
formalization process to other stages of the software lifecycle, i.e.
generating code or test cases from the formal specifications. Not only
does this amortize the cost of creating the specifications, but the
productivity improvement can also be more immediate and easily
measurable.

Driven by the need to shorten and improve the software development
process, Nortel and the Formal Methods Laboratory at the University of
Toronto have jointly proposed a pilot research project to investigate
the feasibility of formal modeling techniques in a commercial
setting. The goal of the project is to find means of using formal
modeling to improve productivity in various stages of the software
lifecycle in an economical manner. Specifically, the emphasis is
placed on deriving test cases from the formal model as the Nortel
engineers have expressed concerns about the feasibility of code
generation for their proprietary platform.

Our exploratory project was organized as a hybrid quantitative and
qualitative case study~\cite{kitchenham96}. As it was extremely
important to choose the right system/language combination for the
formalization process, we began the study by selecting a typical
system to specify and conducting a qualitative evaluation of formal
modeling languages. A chosen language was applied to model the system,
and the resulting model was used to identify errors in the software
requirements document and to derive test suites, shadowing the
existing development process. Throughout the study, we kept a variety
of productivity data for comparison with similar information from the
actual development process. We also noted the qualitative impact of
the formalization process.

The rest of the paper is organized as follows: Section~\ref{sysSelect}
provides a brief description of the software system selected for the
study.  Section~\ref{FMEval} discusses the criteria used in choosing
a suitable modeling language.  In Section~\ref{modeling},
we discuss the formalization process.  Section~\ref{findings} presents
findings from the study. The experience gained during the project is
summarized in Section~\ref{lessonsLearned}.
Section~\ref{presentFutureWork} briefly describes a usability workshop
that we conducted at Nortel, and Section~\ref{conclusions} concludes
the paper.

\section{System Selection}
\label{sysSelect}
\begin{figure}[h]
\begin{center}
\leavevmode
\epsfysize=1.6cm
\epsfbox{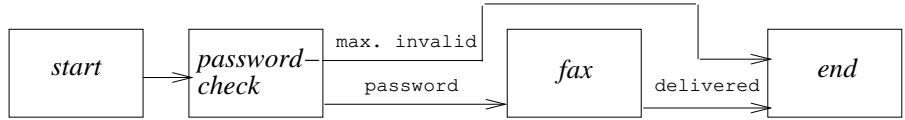}
\end{center}
\vspace{-0.3in}
\caption{\label{figBlocks} A simplified telephony application example.}
\end{figure}

To make the project meaningful, we did not want to be directly
involved in choosing a system, hoping to work on something
representative of typical projects of the TorMAC division of Nortel.
We also felt that it was important to do the formalization in parallel
with the development cycle.  Thus, a group of Nortel engineers,
consisting of developers from the design team and testers from the
verification team decided that we should work on a subsystem of the
Operation, Administration and Maintenance (OAM) software of a
multimedia-messaging system connected to a private branch
exchange. The subsystem, called ServiceCreator in our paper, is a
voice service creation environment that lets administrators build
custom telephony applications in a graphical workspace by connecting
predefined voice-service components together.  Figure~\ref{figBlocks}
illustrates the graphical view of one such telephony application
consisting of four components: {\em start}, {\em end}, {\em
password-check}, and {\em fax}. When the application is
activated, a call session begins at the {\em start} component, and a
caller is required to enter a numerical password in order to retrieve
a fax document from the {\em fax} component. The caller is directed to
the {\em end} component if an incorrect password is entered. In both
scenarios, the call session ends when the {\em end} component is
reached.

\noindent The lines connecting various components represent potential control
flow of the call session, and the actions performed by the caller in an active
component determine the output path taken. In the {\em password-check}
component, for instance, the caller exits via the path {\em password} if a
correct password is entered, or the path {\em max. invalid} if there are too
many invalid password attempts. 

\begin{figure}[h]
\begin{center}
\leavevmode
\epsfbox{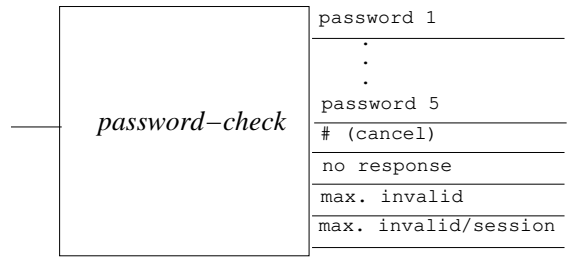}
\end{center}
\vspace{-0.3in}
\caption{\label{figPW} The graphical view of the password-check component.}
\end{figure}

In our study, we analyzed the run-time behavior of 15 such components,
described by an 80-page natural language specification. We illustrate
the approach using the password-check component, described by a 5-page
natural language specification.  Figure~\ref{figPW} shows a graphical
view of this component.

The purpose of this component is to validate digits entered by a
caller against any of the passwords (up to five) defined by the
administrator. For instance, the path {\em password 1} is taken if the
entered digits match the first defined password. The caller is forced
to leave the component using the {\em max. invalid} output if the
maximum number of invalid password attempts is reached. Such attempts
are also monitored on a per-call session basis, and the caller leaves
via the {\em max. invalid/session} output if the per-call session
limit is reached. The caller can also enter the {\em *} key to
retrieve the help prompt, which has a side effect of clearing the
password entry, or the {\em \#} key to exit prematurely via the {\em
\# (cancel)} path if no password has been entered. If the caller stays
idle for a certain time period and has not previously keyed in any
digits prior to entering the password-check component, she is assumed
to be using a rotary phone and is transferred to a different
voice-service component. Otherwise, one of the two delay prompts may
be played depending on whether she has begun keying in the
password. After two more timeouts, the caller exits via the {\em no
response} path.

In order to generalize from the results of the study, we need a good
characterization of the type of applications that ServiceCreator
represents.  First of all, it is clearly a reactive system in the
telecommunications domain.  Additionally, it has relatively stable
requirements and is fairly self-contained, having a loose coupling
with the underlying system.  Finally, it is not very complex although
non-trivial.

\section{Evaluation of Modeling Methods}
\label{FMEval}
A successful formalization of the system in a commercial setting
depends crucially on a modeling language, supported by an appropriate
tool.  Thus, in this section we use the term {\em method} to indicate
both the modeling language and its tool support.  The goal of the
evaluation was to select a suitable modeling method to be used in the
feasibility study.

Easily readable and reviewable artifacts as well as a simple
notation were the two basic requirements for a modeling method to be
usable in a commercial setting.  Moreover, since one of the overall
objectives was to amortize the cost of creating a formal
specification, we began the evaluation by conducting a broad
survey~\cite{andrediary} of available tools that provided support for both
modeling and testing. These constraints turned out to
be extremely limiting as most of the surveyed methods either had just
the modeling or just the testing support, or did not have a formal
notation.  Some were simply too difficult to be used in industry.  We
eventually narrowed down our search to the following candidates:

\BEGINITEM

\item {\em Telelogic SDT}~\cite{telelogic}---an integrated software modeling
and testing tool suite that utilizes Specification and Description Language
(SDL)~\cite{sdl93}, which is based on extended finite state machines (EFSMs),
for behavioral modeling and Message Sequence Charts (MSCs)~\cite{msc93} for
component-interaction specification. MSCs, which can be used as test cases, can
be derived semi-automatically from an SDL model. Alternatively, SDT can verify
the correctness of the model with respect to independently created MSCs.

\item {\em Aonix Validator/Req (V/Q)}~\cite{aonix}---a test generation tool
that allows black-box requirements of a system to be specified in the form
of parameterized UML~\cite{uml97} use cases. Validator/Req generates test cases
for data-flow testing from the model automatically and produces documents that
conform with the IEEE standard for software requirements
specifications~\cite{ieee830}. 

\item {\em Teradyne TestMaster (TM)}~\cite{teradyne}---a test generation tool
that automatically generates test cases for both control-flow and limited
data-flow testing from models based on EFSMs. The number of test cases can be
flexibly tuned to suit different testing needs.

\ENDITEM

To perform a detailed assessment, we structured our evaluation as a feature
analysis exercise~\cite{kitchenham96} and refined our focus to choosing among
the remaining methods using additional evaluation criteria gathered from the
Nortel engineers. These criteria comprised of factors, such as usability and
smooth integration, that were crucial to the use of formal modeling in their
environment. After the methods were used to model the password-check component,
they were ranked against the criteria based on our impressions of the tool
and the models produced.

\begin{table}[h]
\centering
\begin{footnotesize}
\begin{tabular}{|p{3in}|c||c|c|c|} \hline 
{\bf Criteria} & {\bf Weighting} & {\bf SDT} &
{\bf V/Q} & {\bf TM} \\ \hline \hline

has modeling support                        & 5 & 4 & 2 & 3\\ \hline
has testing support                         & 5 & 3 & 4 & 5\\ \hline
has a gentle learning curve                 & 4 & 3 & 5 & 3\\ \hline
produces easy-to-understand artifacts       & 3 & 4 & 5 & 4\\ \hline
enhances understanding of the system        & 4 & 5 & 3 & 4\\ \hline
works smoothly with the existing development process and tools
                                            & 4 & 5 & 4 & 3\\ \hline
is scalable                                 & 3 & 4 & 2 & 3\\ \hline
has strong tool support               & 3 & 5 & 4 & 5\\ \hline
performs completeness or consistency checks & 2 & 4 & 3 & 3\\ \hline
provides features such as multi-user support, requirements traceability and
document generation                         & 1 & 3 & 4 & 2\\ \hline \hline
{\bf Score} & N/A & {\bf 137} & {\bf 121} & {\bf 124} \\ \hline

\end{tabular}
\end{footnotesize}
\caption{\label{tableEval} Comparison of modeling methods.}
\end{table}

Table~\ref{tableEval} shows results of this evaluation. It lists the
evaluation criteria in column one, their relative importance using a
scale from 1 (least important) to 5 (most important) in column
two\footnote{These were assigned after consultations with Nortel
engineers.}, and the degree the methods satisfy the criteria using a
scale from 1 (unsatisfactory) to 5 (completely satisfactory) in
columns three to five. In the first row, for instance, SDL scores the
highest as it allows the behavioral modeling and the hierarchical
partitioning of a system into concurrent processes. TestMaster has
similar modeling support but a system can only be modeled as a single
hierarchical EFSM from a tester's perspective. Validator/Req ranks the
lowest as its scenario-based notation provides very limited modeling
support. The conclusion of the evaluation was reached by comparing the
final scores for the methods obtained by adding weighted scores from
each criterion, making SDT the most suitable tool.

A confounding factor in feature analysis is the potentially biased
opinions of the evaluator. Although we tried to ensure the objectivity
of the evaluation, assignments of the scores inevitably contained our
subjective judgment, and we did not feel that we could accurately
evaluate such factors as usability. To mitigate this potential problem
and to gain more confidence in our assessment, we demonstrated the
tools and the models to the engineers. They agreed that SDT satisfied
their criteria more closely than the other tools.

\section{Modeling and Testing the System}
\label{modeling}
\begin{figure}[h]
\begin{center}
\leavevmode
\epsfysize=7cm
\epsfbox{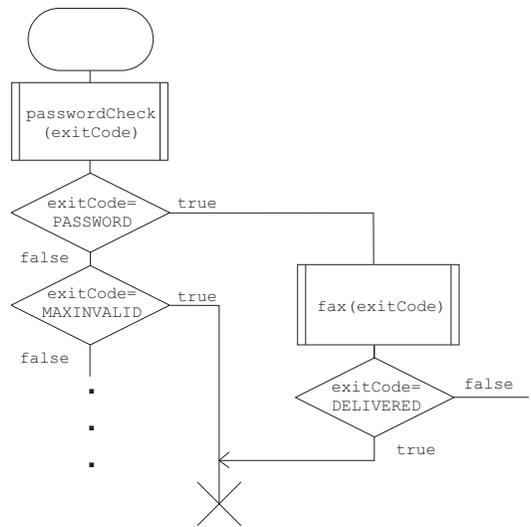}
\end{center}
\vspace{-0.3in}
\caption{\label{driver} A driver process for a simplified telephony application example of Fig.~\ref{figBlocks}.}
\end{figure}

The next step was to formalize the ServiceCreator application.  This
formalization was undertaken by us in parallel with the actual
development process. ServiceCreator was modeled as a 70-page SDL
system in which the environment contained the underlying OAM software
and messaging-system hardware. Out of the 15 voice-service components,
10 were modeled as separate SDL processes (see Appendix A for an
example) that communicated with the environment through a ``driver''
SDL process.  This process models the control-flow information of the
telephone application.  Figure~\ref{driver} shows a simplified view of
the ``driver'' process created for the application of
Figure~\ref{figBlocks}.  This process is responsible for activating
voice-service components and responding to their termination.
The functionality of the five remaining components (e.g., {\em start}
and {\em end}) was incorporated into the ``driver'' processes. A
total of 23 signals were used in the SDL system; eight of them were
external (used for communicating with the environment) and the rest
internal. Persistent data such as the predefined passwords for the
password-check component, were represented as parameters to the SDL
processes.

\subsection{The Level of Abstraction}
\label{abstraction} 
The modeling process was relatively straightforward as we encountered
no major problems in modeling ServiceCreator; we also felt that a
background in formal methods was not required. However, the biggest
concern was to determine the appropriate level of abstraction which
was dictated by two opposing needs: the model should be constructed
from a black-box view to reduce its complexity, while the exact
behavior of the system needs to be modeled for deriving detailed test
cases. In addition, a more detailed model would help in identifying
problems in the natural-language specification.

\begin{figure}[h]
\begin{center}
\leavevmode
\epsfbox{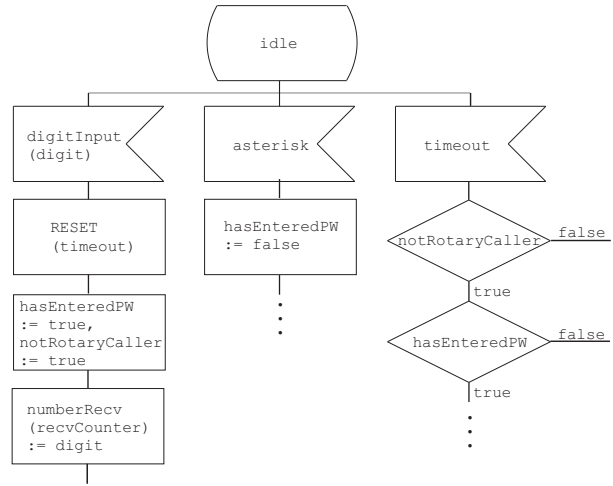}
\end{center}
\vspace{-0.3in}
\caption{\label{figSDL} A simplified fragment from the SDL model of the
password-check component.}
\end{figure}

Our approach was to start from a high level of abstraction, filling in
details about some parts of the behavior if the natural-language
specification required it. Mostly, such details were necessary in
dealing with external input.  For example, in modeling the
password-check component, we represented the various timeouts by an
SDL timer {\em timeout}~(see Figure~\ref{figSDL}), as the actual
length of the timeouts was relatively unimportant. Processing of user
input, on the other hand, required reasoning on the level of single
digits. In our model, a received digit, {\em digit}, was actually
stored in a variable {\em numberRecv}. While this treatment could
potentially lead to large and cluttered models, we sometimes had to
resort to it to be able to derive sufficiently detailed test cases.

\subsection{Test Case Derivation}
\label{testCase}
To obtain immediate benefits from the formalization process, 120 MSCs
were derived from the SDL model for testing the implementation. The
derivation was not automatic: SDT recorded the interactions between
the SDL system and its environment as MSCs when we animated the SDL
model manually.  We felt that the automation was not necessarily
desirable since this exercise gave us confidence in the content and
the coverage of the test cases.

During the test case derivation, we took advantage of the modular
nature of the voice-service components and generated test cases for
each of them separately, achieving full transition coverage in the
corresponding SDL processes.  However, some functionality of the
system could be covered only by testing multiple components together, i.e.
by integration testing.  More than 20 integration test cases
have been identified.  
For instance, to derive test cases for testing the initial timeout
behavior for touch-tone callers where no passwords were keyed in, we
created a telephony application model in which a caller was required
to key in some digits in a component, such as a menu, prior to
entering the password-check component (see Appendix B for the
corresponding MSC). The procedure for deriving such test cases
was as follows:

\BEGINITEM

\item During the modeling phase, note the cases where the input comes not
only from the environment but also from other components.  If some input
to component $A$ comes from component $B$, we say that there is
a relationship between $A$ and $B$.
\item After the modeling phase, use the resulting model to create
test cases that specifically ensure that the relationship between
the components is correctly implemented.

\ENDITEM

Derivation of test cases for integration testing was the most 
labor-intensive part of this phase; it also required a fair bit
of skill.

\subsection{Specification and Implementation Problems}
\label{errors}
The SDL model and the experience gained from the formalization process
allowed us to identify specification errors that escaped a series of
manual inspections by Nortel engineers. As the components modeled were
not particularly complicated, most of the errors we found were caused
by vagueness and missing information. In fact, the most time-consuming
part of the formalization process was to understand the
natural-language specifications and to consult the engineers for
clarifications. We estimated that these activities took as much time
as the formalization process. Some of the problems found in the
specification of the password-check component are presented below:

\setcounter{numList-count}{1}
\BEGINNUMITEM

\item it was never mentioned whether the various voice prompts were
interruptible, and no conventions for such behavior were defined;

\item lengths of various timeouts were not mentioned;

\item it was unclear which exit path should be taken when the administrator set
two or more passwords to be the same;

\item the maximum and minimum lengths of passwords were not defined.

\ENDNUMITEM

\noindent Some of these problems were very low criticality and could 
be easily fixed in the implementation.  However, the testers were
required to interact with the developers to clarify the correct
behavior of the system, spending an unnecessarily long time in the
test case execution phase (see Section~\ref{quanresult}).  In
addition, problem 4) propagated itself into the implementation.  That
is,
a malicious caller could crash the system by entering an abnormally
long password within the password-check component.  Thus, this
requirement omission became a critical implementation error.
Moreover, since the MSCs derived from the SDL model were used to
identify errors in the implementation {\em after} the Nortel engineers
had officially completed testing of the voice-service components, we
were able to observe that this implementation error had not been
discovered.  The reason was that the test cases were derived from the
same incomplete specification and missed a critical error.



\section{Findings}
\label{findings}
We began our analysis by seeking quantitative evidence to measure the
effects of the formalization process on productivity. However, as the
study progressed, we felt that it was also crucial to identify the
qualitative factors (e.g., perceived usability of SDL and commitments
from the development team) and the limitations of the study in order
to reach accurate and unbiased conclusions.

Unfortunately, Nortel engineers did not keep track of many essential
metrics and, due to the exploratory nature of our study, did not allow us
to create a controlled environment where such metrics could be
obtained.  In particular, we do not know the exact amount of time it
takes to fix a bug, if it is found during the inspection vs.  design
vs. testing phase.  The lack of metrics significantly impairs our
ability to draw quantitative conclusions.

\subsection{Quantitative Results}
\label{quanresult}
The entire modeling process, which consisted of activities such as
understanding and formalizing the specification as well as deriving
and executing test cases, took about two person-months to
complete. During this period, we kept track of a variety of
productivity data in the study (column two of Table~\ref{tableData})
for comparison with similar information from the existing development
process (column three). Effort measurements in this table are
approximated to the nearest person-day.

\begin{table}[h]
\centering
\begin{footnotesize}
\begin{tabular}{|l||c|c|} \hline
{\bf Productivity Data} & {\bf Study} & {\bf Existing Process}
\\ \hline \hline

Time to model (person-days)                &  11 & N/A \\ \hline
Time to inspect (person-days)              & N/A &  50 \\ \hline
Number of specification errors reported    &  56 & N/A \\ \hline
Number of test units                       & 269 &  96 \\ \hline
Time to create test units (person-days)    &   7 &   7 \\ \hline
Time to execute test units (person-days)   &   5 &  14 \\ \hline
Number of implementation errors identified &  50 &  23 \\ \hline

\end{tabular} 
\end{footnotesize}
\caption{\label{tableData} Productivity Comparisons.}
\end{table}

\noindent Since the sizes of test cases varied greatly, we did not use a ``test
case'' as the unit of comparison for testing-related data. Instead, we
counted test units, the smallest externally visible functionality of
the system, in each test case to ensure a fair comparison. Highlights
from the table are summarized below:

\BEGINITEM

\item The {\em time to model} value included only the time used for modeling
the SDL processes. While the modeling task did not have an equivalent
in the actual development process, manual inspection was a similar
activity that was also performed at the completion of the
specification phase. Certainly, the formalization process was not meant
to be a complete replacement. However, if a large number of
specification errors were identified in a relatively short amount of
time, the modeling task could be considered a way to decrease the
time for inspection.  (We discuss this point later in this Section.)

\item The number of specification errors reported could not be used for
comparison as the Nortel engineers did not keep track of such
statistics.

\item More test units could be derived from the SDL model (which translated to
better test coverage) in roughly the same amount of time, possibly
because the model eased the creation task by providing a more in-depth
understanding of the system as well as a better sense of
completeness. One other reason for the difference in the quantity was
that the test units from Nortel were sometimes vaguely specified (see
Figure~\ref{figNortelTC} for an example); the missing details
contributed to a decrease in the number of test units reported.

\begin{figure}[h]
\centering
\fbox {\parbox{4in}{
Call the application and simulate a number of hacker attempts (such as
common passwords and misspelled passwords). Verify that the
application terminates the call session after the maximum number of
password attempts is reached.  } }
\caption{\label{figNortelTC} A Nortel test case for testing the password-check
component.}
\end{figure}

\item The time needed for test unit execution in our study was much 
smaller for two major reasons: the derived test units were more
detailed and thus easier to execute, and it was observed that Nortel
engineers spent a lot of time revising the existing test cases
because of the changes in requirements and creating more detailed test
scenarios based on the vaguely specified test units. However, due to
tight schedules, most testers did not document these extra units until
the end of the entire testing phase, which spanned over more than four
months. They admitted that some of these test units would inevitably
be lost, contributing to a decrease in their total number.

\item The number of implementation errors identified in the study was 
two times larger than that of the existing development process. Many
of them were missed because testers created test cases from an
incorrect and incomplete specification, as indicated in the third row
of Table~\ref{tableData}. Problems such as incompleteness propagate to
the test cases and affect the test coverage. In fact, 18 of the 50
problems could be linked to problems in the specification.  Their
criticality ranged from the minor ones, dealing with the availability
and the interruptability of voice prompts, to the critical ones,
affecting the core functionality of the voice-service components or
causing the system to crush.  Most of these errors resulted from
undocumented assumptions or incorrect/missing error handling.
\ENDITEM

To obtain an accurate cost/benefit figure, we needed to
collect additional statistics such as an average cost to fix a
requirements error discovered during the inspection and the
implementation phases.  As we mentioned above, the Nortel development
team did not keep track of such metric; however, a conservative
cost/benefit estimation is still possible.  Without taking the
improvement in software quality into account, we can estimate the
cost of formalization by subtracting the time of the modeling task
from the direct savings in

\BEGINITEM
\item the test unit creation (0 person-days),
\item the test unit execution (9 person-days), and
\item the manual inspection.
\ENDITEM

\noindent The formalization process did not include a manual 
inspection phase, whereas the actual development took 50 person-days
for it.  The modeling task would come at no cost if it were to reduce
the manual inspection by 2 person-days, or 4\%.  Of course, the actual
cost/benefit figure is significantly more promising if the long-term
benefits, coming from a better quality of the product, ease of
maintenance and regression testing, and an ability to reuse a good
specification, are taken into account.

\subsection{Qualitative Observations}
\label{qualobserve}
While all the quantitative data was in favor of the use of the formal
modeling, it was clear that these results alone constituted only a
part of our findings.  Some observations from the formalization process 
that were not evident from Table~\ref{tableData} are discussed below.

\BEGINITEM

\item The most frequent complaint from the test engineers is that the 
missing information in the specifications often complicates the task
of the test case creation. The SDL models encourage and assist
developers in stating system requirements more precisely and
completely, which should allow testers to create better quality (e.g.,
more detailed and with expected results more clearly defined) test
cases and reduce the time needed for test case creation and execution.
Developers should also benefit from the more complete specifications
during the design and the implementation phases. This is an area where
SDL can potentially significantly improve the development process. In
fact, SDL has been successfully applied in the telecommunications
field: from the traditional use of protocol
specifications~\cite{belina91,sarma91,facchi96} to high level
specification~\cite{mansson93}, prototyping~\cite{vogel98},
design~\cite{kooij98}, code generation~\cite{froberg93}, and
testing~\cite{grasmanis91} of telecommunications applications.
Although the results reported in these studies were similar to ours,
the goal of the studies was different: they were aimed to investigate
technical advantages or a feasibility of SDL in a given environment,
or were emphasizing only one of the development activities.



\item As with any other formal specification technique, a successful
integration of SDL into the development process requires a firm
commitment from the entire development team. For instance, the
developers must ensure that the SDL model is always kept consistent
with the system requirements and the code, e.g. last-minute changes in
the design and implementation are propagated back to the
model. Testers also need to ensure that their test cases always
reflect the model accurately. While this is possibly one of the
biggest obstacles in applying a formal modeling technique, the
advantages provided by SDL justify the extra effort.

\item Compared to other formal modeling techniques, the strengths of 
SDL lie in its ease of use and the ability to express nontrivial
behavior in a reviewable notation. Unlike many other formal modeling
languages, SDL does not require an explicit use of formal logic.  The
graphical user-friendly notation allows developers without a strong
mathematical background to effectively create EFSMs.  Compared to
natural language specifications, such EFSMs give a much better sense
of completeness, allowing to easily detect missing scenarios, e.g.,
problem 1) in Section~\ref{errors}.  In addition, the formal syntax helps
clarify ambiguities or inconsistencies, e.g., problem 3) in the same
section. However, SDL tends to blur the line between requirements and
designs. If proper abstractions are not applied, the model may become
too detailed and unnecessarily large, possibly duplicating the design
effort. 

\ENDITEM

\subsection{Limitations}
\label{limitations}
Based on the opinions expressed by the Nortel developers and testers,
ServiceCreator was representative of the types of systems they
have to work with, so we are fairly confident that the results of the
study would apply to similar projects in this environment.  We were
also fortunate to find a method which is well suited for modeling
telecommunication systems.  However, it would be
difficult~\cite{zelkowitz98} to generalize our findings outside the
Nortel domain, since they would depend on the current development
methodology, types of applications and the choice of a modeling
language/tool.


Other limitations came from the fact that we had prior experience with
SDL and were not constrained by development pressures.  That is, we
took the time necessary to produce high quality models and detailed
test cases and felt that the process was straightforward.  If time
pressures prevent Nortel developers from applying the modeling
techniques carefully, they may not achieve equally good results.  In
addition, novice users of SDL would take more time and possibly create
less effective models of their systems.  However, we believe that
appropriate training and availability of an SDL expert can ensure that
Nortel engineers use the SDL system successfully.


\section{Lessons Learned}
\label{lessonsLearned}

We were able to show that formal modeling techniques can shorten the
development cycle while improving the quality of software.  This can
be done by amortizing the cost of creating the model over
time-intensive phases of the software development lifecycle, such as
testing or code generation.  However, the total decrease in the
development cycle is only achievable if the formalization can be done
fairly inexpensively, by utilizing an easy-to-use and review notation,
formalizing only selected components, and staying at a fairly high
level of abstraction.  It is also essential to achieve immediate
results by using the approach incrementally, that is, being able to
stop at any time and get partial benefits from partial modeling.  A
light-weight approach to formalization has been advocated by many
researchers~\cite{davis96,jackson96b,jones96} and applied successfully
in several projects, e.g.~\cite{easterbrook98a,feather98}.

What about verification?  We feel that in the current commercial
environment the majority of systems do not require any verification.
There is typically a lesser need for absolute assurance, but a greater
need for rapid development of reasonably correct systems.  In fact,
our use of SDL showed that, if verification is not involved, it is not
essential to use a modeling language with a fully-defined formal
semantics to achieve immediate and measurable benefits.

\section{Measuring Usability of SDL}
\label{presentFutureWork}

There is no doubt that usability of formal modeling techniques plays
an important role in their acceptance in
industry~\cite{heitmeyer98a}. An easy-to-use technique
encourages experimentation and reduces the cost of integration. More
importantly, the reality is that practitioners do not try to adapt to an
inconvenient technique---they simply abandon it~\cite{weber93}. Thus,
it is essential that SDL is perceived to be usable by Nortel
engineers.  Only then will they be willing to apply it to their projects.

To collect some information about the usability of SDL, we conducted a
one-day workshop in which six Nortel engineers participated.  In the
first part of the workshop we provided the participants with 
natural language descriptions of two small software
systems~\cite{wong99}. After inspecting the descriptions manually and
noting problems in them, the participants were asked to model the
described systems in SDL. By formalizing the behavior, they were able
to discover many additional specification errors; some of them found
even more errors than we originally seeded, i.e.  the descriptions
contained errors that we did not notice. A few minor usability
problems were noted, but the consensus reached among the participants
was that the use of a formal, yet user-friendly, notation could help
uncover problems hidden in the seemingly simple exercises much more
effectively than manual inspections. In the second part of the
workshop we asked the participants to fill in a questionnaire.  The
goal of the questionnaire was to obtain opinions about the usability
of SDL and its perceived role in the development environment.  Some of
the results are summarized in Table~\ref{tableQuest}, and the rest are
available in~\cite{wong99}.  The column on the right contains an
average score on the scale from 1 (strongly disagree) to 5 (strongly
agree).

\begin{table}[h]
\centering
\begin{footnotesize}
\begin{tabular}{|p{3.5in}||c|} \hline
\begin{centering} {\bf Statement} \end{centering} & {\bf Score}
\\ \hline \hline

SDL is easy to use
& 3.7 \\ \hline
SDL can be used to address many of the development problems we are facing.
& 3.7 \\ \hline
The use of SDL increases our understanding of the requirements and their quality.
& 4.2 \\ \hline
The use of SDL lengthens the requirements analysis phase.
& 4.0 \\ \hline
The use of SDL shortens the design, implementation, and testing phases.
& 4.1 \\ \hline
Integrating SDL into my work routine is worthwhile and should be tried.
& 3.7 \\ \hline
\end{tabular} 
\end{footnotesize}
\caption{\label{tableQuest} Some results from the questionnaire.}
\end{table}

Results from the questionnaire strengthened our findings that SDL is a
user-friendly formal modeling technique which can be used effectively
by Nortel engineers to improve their development process. Encouraged
by the prospects of SDL, Nortel and University of Toronto are in the
process of setting up another joint project where the engineers will
carry out the formalization process themselves, and we will only
observe the progress and provide consulting, if necessary. Without
many limitations of our study, this new project will provide a more
accurate insight into the technical and economical feasibility of SDL
in a commercial setting.


\section{Conclusions}
\label{conclusions}
In this case study we formalized the behavior of a
multimedia-messaging system in a commercial setting.  The success of
the study was in finding a representative system, carefully selecting
a suitable modeling method, and taking a lightweight formalization
approach.  Although we did not have access to some development metrics
to fully quantify our findings, the results of the study clearly show
that software requirements can be formalized effectively and
economically, yielding significant improvements over the existing
development process.

\vspace{6mm}

\noindent
{\large\bf Acknowledgments}

\vspace{4mm}

\noindent The authors would like to thank Albert Loe, Steve Okun, Avinash Persaud, and
Shawn Turner of Nortel for their technical assistance and continual
support throughout the study. We are also grateful to the anonymous
referees for suggesting ways to improve the paper.  The study was
supported in part by Nortel Networks, NSERC, and Ontario Graduate
Scholarship (OGS).

\newpage

\appendix

\section{An SDL Block Diagram}
The diagram below illustrates the SDL block diagram of the telephony
application shown in Figure~\ref{figBlocks}. The ``driver'' block (which
contains a ``driver'' process) acts as the signal router between the
environment and the SDL blocks of the voice-service components by routing the
signal lists {\em inputSigLst} and {\em outputSigLst}. It is also responsible
for activating the appropriate component according to the control-flow of the
application and actions from the caller by using the signal {\em activate}.
Please refer to Section~\ref{modeling} for more details.

\begin{figure}[h]
\begin{center}
\leavevmode
\epsfysize=6cm
\epsfbox{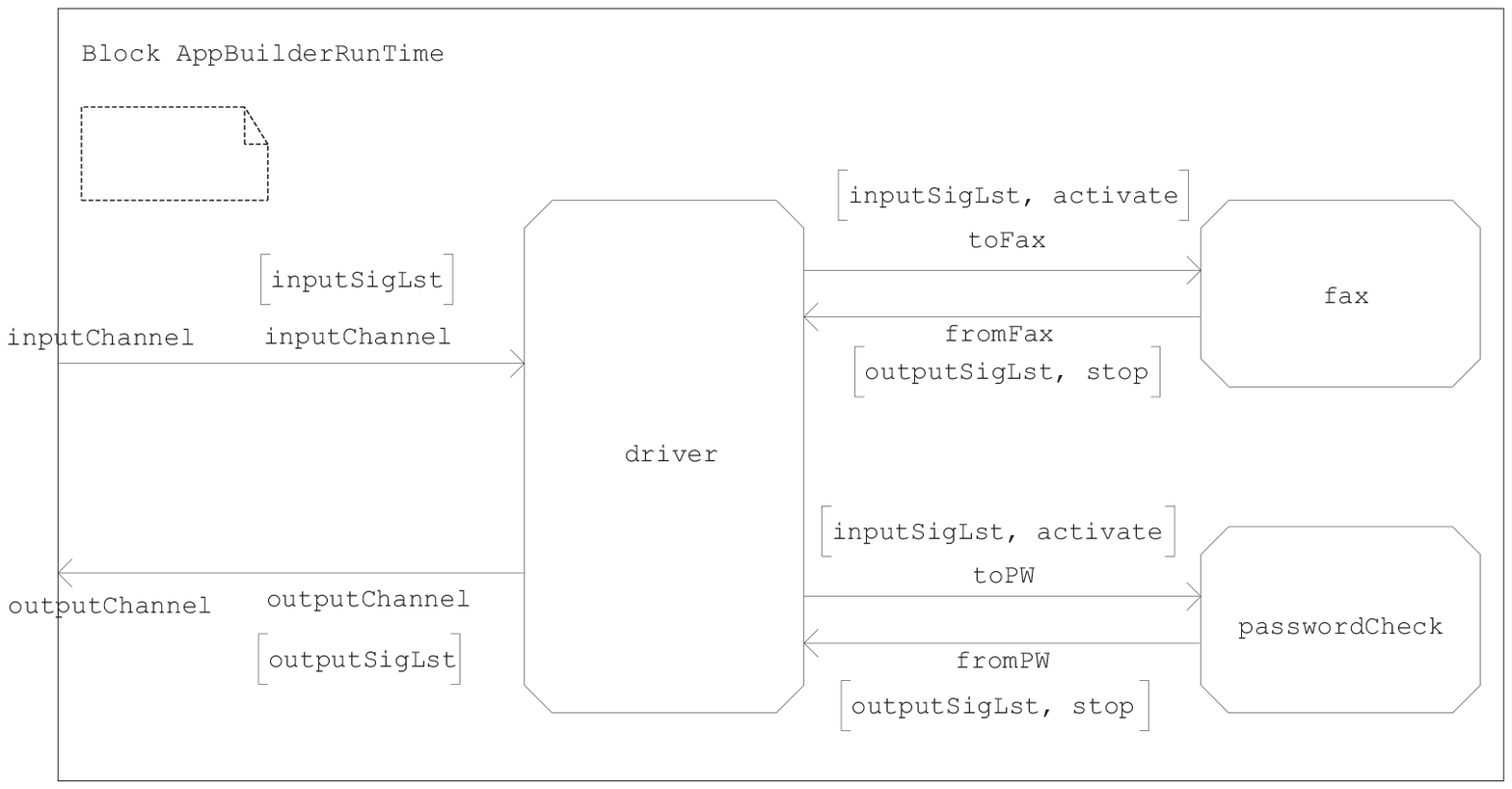}
\end{center} \vspace{-0.3in}
\end{figure}

\newpage

\section{A Message Sequence Chart}

The MSC below shows the interactions between a caller and a telephony
application in ServiceCreator. The application requires a caller
to press button {\em 1} in the menu component prior to entering the
password-check component. Refer to Section~\ref{testCase} for more
details.

\begin{figure}[h]
\begin{center}
\leavevmode
\epsfysize=9.5cm
\epsfbox{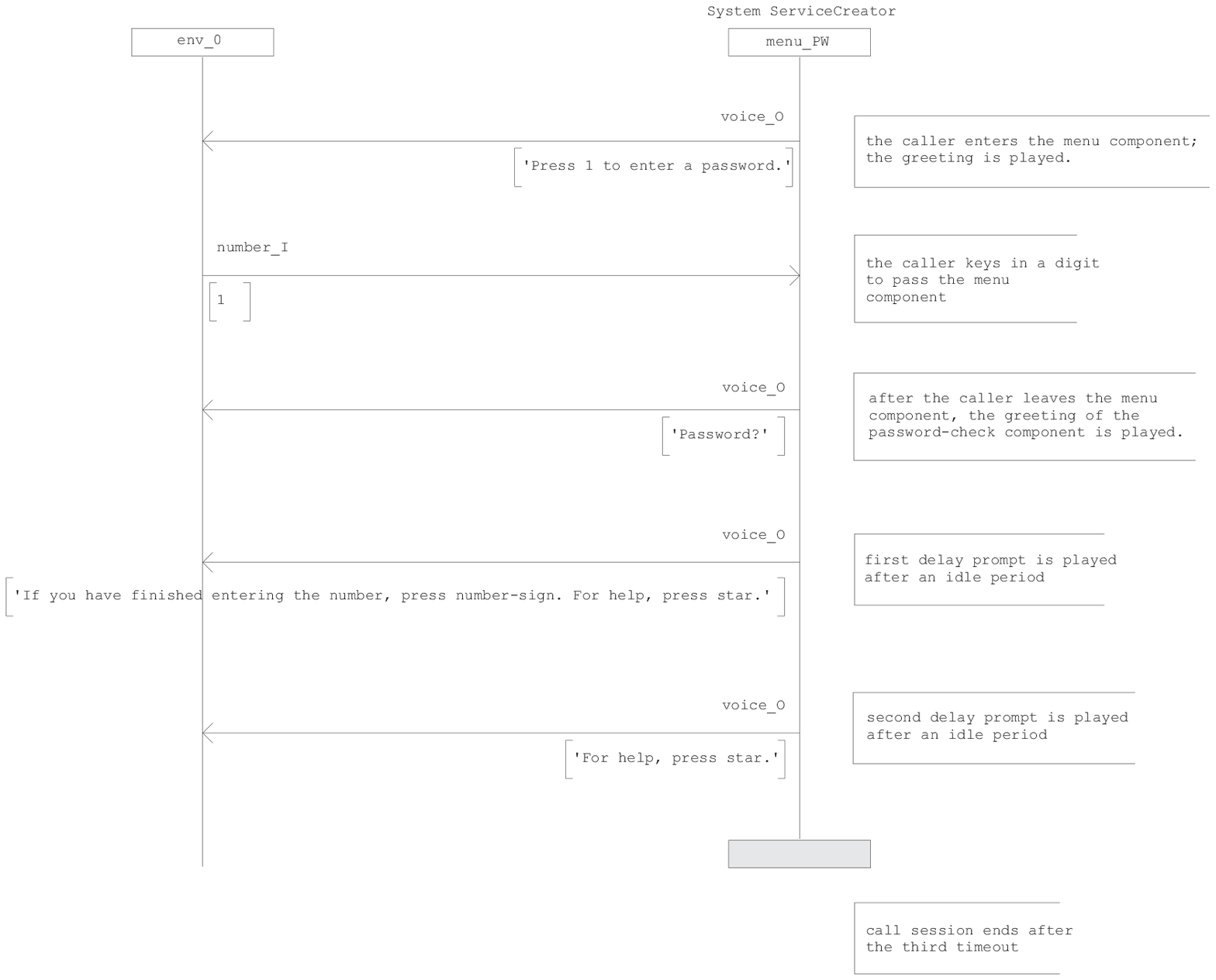}
\end{center} \vspace{-0.3in}
\end{figure}


\newpage

\begin{thebibliography}{10}

\bibitem{ieee830}
ANSI/IEEE.
\newblock {``IEEE 830: IEEE Recommended Practice for Software Requirements
  Specifications ''}.
\newblock {\em IEEE}, 1993.

\bibitem{aonix}
Aonix.
\newblock {``Aonix Home Page''}.
\newblock \url{http://www.aonix.com}, September 1998.

\bibitem{belina91}
F.~Belina, D.~Hogrefe, and A.~Sarma.
\newblock {\em SDL with Applications from Protocol Specification}.
\newblock Prentice Hall, 1991.

\bibitem{uml97}
UML~Partners Consortium.
\newblock {``UML Proposal Document Set''}.
\newblock {\em OMG documents ad/97-08-\{02,03,04,05,06,07,08,09\}}, September
  1997.

\bibitem{crow96}
J.~Crow and B.L.~Di Vito.
\newblock {``Formalizing Space Shuttle Software Requirements''}.
\newblock In {\em Workshop on Formal Methods in Software Practice}, San Diego,
  California, January 1996.

\bibitem{davis96}
R.E. Davis and R.L. Danielson.
\newblock {``Practically Formal Methods''}.
\newblock In {\em Proceedings of International Conference on Software
  Engineering: Education and Practices}, pages 168--175. IEEE Computer Society
  Press, January 1996.

\bibitem{easterbrook98a}
Steve Easterbrook, Robyn Lutz, Richard Covington, John Kelly, Yoko Ampo, and
  David Hamilton.
\newblock {``Experience Using Lightweight Formal Methods for Requirements
  Modeling''}.
\newblock {\em IEEE Transactions on Software Engineering}, 24(1):4--14, January
  1998.

\bibitem{facchi96}
Christian Facchi, Markus Haubner, and Ursula Hinkel.
\newblock The SDL Specification of the Sliding Window Protocol Revisited.
\newblock Technical Report TUM-I9614, Technische Univerit\"at M\"unchen, 1996.

\bibitem{feather98}
M.S. Feather.
\newblock {``Rapid Application of Lightweight Formal Methods for Consistency
  Analysis''}.
\newblock {\em IEEE Transactions on Software Engineering}, November 1998.

\bibitem{froberg93}
M.W. Froberg.
\newblock {``Automatic Code Generation from SDL to a Declarative Programming
  Language''}.
\newblock In {\em Proceedings of the Sixth SDL Forum}, Darmstadt, Germany,
  October 1993.

\bibitem{grasmanis91}
M.~Grasmanis and I.~Medvedis.
\newblock {``Approach to Behaviour Specification and Automated Test Generation
  for Telephone Exchange Systems''}.
\newblock In {\em Proceedings of the Fifth SDL Forum}, Glasgow, Scotland,
  September 1991.

\bibitem{hall96}
Anthony Hall.
\newblock {``Using Formal Methods to Develop an ATC Information System''}.
\newblock {\em IEEE Software}, 13(2), March 1996.

\bibitem{heimdahl96a}
Mats~P.E. Heimdahl.
\newblock {``Lessons from the Analysis of TCAS II''}.
\newblock In {\em Proceedings of the International Symposium on Software
  Testing and Analysis (ISSTA'96)}, San Diego, CA, January 1996.

\bibitem{heitmeyer98a}
Constance Heitmeyer.
\newblock {``One the Need for Practical Formal Methods''}.
\newblock In {\em Proceedings of 5th Int. Symp. on Real-time and Real-time
  Fault Tolerant Systems (FTRTFT'98)}, pages 18--26, 1998.
\newblock LICS 1486.

\bibitem{hoare95}
Jonathan~P. Hoare.
\newblock {``Application of the B-Method to CICS''}.
\newblock In M.~G. Hinchey and J.~P. Bowen, editors, {\em Applications of
  Formal Methods}, pages 97--124. Prentice Hall International Series in
  Computer Science, 1995.

\bibitem{sdl93}
ITU-T.
\newblock {``ITU-T Recommendation Z.100: Specification and Description Language
  (SDL)''}.
\newblock {\em ITU-T}, 1993.

\bibitem{msc93}
ITU-T.
\newblock {``ITU-T Recommendation Z.120: Message Sequence Chart (MSC)''}.
\newblock {\em ITU-T}, 1993.

\bibitem{jackson96b}
Daniel Jackson and Jeannette Wing.
\newblock {``Lighweight Formal Methods''}.
\newblock {\em IEEE Computer}, April 1996.

\bibitem{joannou93}
Paul~K. Joannou.
\newblock {``Experiences from Application of Digital Systems in Nuclear Power
  Plants''}.
\newblock In {\em Proceedings of the Digital Systems Reliability and Nuclear
  Safety Workshop}, Rockville, Maryland, 1993.

\bibitem{jones96}
Cliff~B. Jones.
\newblock {``An Invitation to Formal Methods: A Rigorous Approach to Formal
  Methods''}.
\newblock {\em IEEE Computer}, 20(4):19, April 1996.

\bibitem{kitchenham96}
Barbara~Ann Kitchenham.
\newblock {``Evaluating Software Engineering Methods and Tools. Part 1''}.
\newblock {\em ACM SIGSOFT Software Engineering Notes}, 21(1):11--15, January
  1996.

\bibitem{mansson93}
L.~Mansson.
\newblock {``High Level Specification of a Telecom Application with SDL'92''}.
\newblock In {\em Proceedings of the Sixth SDL Forum}, Darmstadt, Germany,
  October 1993.

\bibitem{mataga95}
Peter Mataga and Pamela Zave.
\newblock {``Multiparadigm Specification of an AT\&T Switching System''}.
\newblock In M.~G. Hinchey and J.~P. Bowen, editors, {\em Applications of
  Formal Methods}, pages 375--398. Prentice Hall International Series in
  Computer Science, 1995.

\bibitem{kooij98}
M.Kooij and L.~Provoost.
\newblock {``Industrial Report on the Use of Abstraction in SDL/MSC''}.
\newblock In {\em First Workshop of the SDL Forum Society on SDL and MSC},
  Berlin, Germany, June 1998.
\newblock alcatel.

\bibitem{parnas93a}
D.L. Parnas.
\newblock {``Some Theorems We Should Prove''}.
\newblock In {\em Proceedings of 1993 International Conference on HOL Theorem
  Proving and Its Applications}, Vancouver, BC, August 1993.

\bibitem{sarma91}
A.~Sarma.
\newblock {``Modelling Broadband ISDN Protocols with SDL''}.
\newblock In {\em Proceedings of the Fifth SDL Forum}, Glasgow, Scotland,
  September 1989.

\bibitem{telelogic}
Telelogic.
\newblock {``Telelogic SDT Home Page''}.
\newblock \url{http://www.telelogic.com/solution/tools/sdt.asp}, September
  1998.

\bibitem{teradyne}
Teradyne.
\newblock {``TestMaster Home Page''}.
\newblock \url{http://www.teradyne.com/prods/sst/ssthome.html}, September 1998.

\bibitem{vogel98}
H.J. Vögel, W.~Kellerer, S.~Karg, M.~Kober, A.~Beckert, and G.~Einfalt.
\newblock {``SDL based prototyping of ISDN-DECT-PBX switching software''}.
\newblock In {\em First Workshop of the SDL Forum Society on SDL and MSC},
  Berlin, Germany, June 1998.

\bibitem{weber93}
Debora Weber-Wulff.
\newblock {``Selling Formal Methods to Industry''}.
\newblock In J.C.P. Woodcock and P.G. Larsen, editors, {\em Proceedings of
  FME'93: Industrial Benefit of Formal Methods, First International Symposium
  of Formal Methods Europe}, pages 671--678, Odense, Denmark, April 1993.
  Springer-Verlag.

\bibitem{andrediary}
Andre Wong.
\newblock {``The Diary of the Formal-Method Survey''}.
\newblock \url{http://www.cs.toronto.edu/~andre/progress.html}, September 1998.

\bibitem{wong99}
Andre Wong.
\newblock {\em {``Formalizing Requirements in a Commercial Setting: A Case
  Study''}}.
\newblock M.Sc. thesis, University of Toronto, Department of Computer Science,
  Toronto, ON, Canada, 1999.
\newblock (In preparation).

\bibitem{zelkowitz98}
Marvin~V. Zelkowitz and Dolores~R. Wallace.
\newblock {``Experimental Models for Validating Technology''}.
\newblock {\em IEEE Computer}, 31(5), May 1998.

\end{thebibliography}
\end{document}